\documentclass[final,3p,times]{elsarticle}

\biboptions{numbers}

\usepackage{hyperref}
\usepackage{cancel}
\usepackage{graphicx}%
\usepackage{multirow}%
\usepackage{amsmath,amssymb,amsfonts}%
\usepackage{amsthm}%
\usepackage{mathrsfs}%
\usepackage[title]{appendix}%
\usepackage{xcolor}%
\usepackage{textcomp}%
\usepackage{manyfoot}%
\usepackage{booktabs}%
\usepackage{algorithm}%
\usepackage{algorithmicx}%
\usepackage{algpseudocode}%
\usepackage{listings}%

\usepackage{amsfonts}
\usepackage{graphics}
\usepackage{bm}
\usepackage{bbm}
\usepackage[section]{placeins}
\usepackage{caption}
\usepackage{subcaption}
\usepackage{epsfig}
\usepackage{booktabs}
\usepackage{color}
\usepackage{mathtools}
\graphicspath{ {./figures/} }
\usepackage{booktabs}
\usepackage{multirow}
\usepackage{pifont}
\usepackage{inputenc}
\usepackage{tikz}
\usetikzlibrary{shapes,arrows}
\usetikzlibrary{shapes.geometric}
\usetikzlibrary{calc}

\usepackage{lineno}
\usepackage{mathrsfs}
\usepackage{color}
\usepackage{xcolor}

\journal{Elsevier}

\begin{document}

\begin{frontmatter}



\title{Phase-Field Fracture Simulation of Highly Deformable Thin Structures via a Discrete Differential Geometry Framework}

\author{Bohan Zhang $^{a,b}$, Bo Wang $^{a,c,}$\footnote[1]{\textit{Corresponding Author: bowang@nwpu.edu.cn (B.W.)}}, Huajiang Ouyang $^{d}$,
Zhigang Wu $^{e}$,
Yu Zhou $^{f,}$\footnote[1]{\textit{Corresponding Author:  yuzhou@seas.harvard.edu (Y.Z.)}}, Weicheng Huang $^{b,}$\footnote[1]{\textit{Corresponding Author: weicheng.huang@newcastle.ac.uk (W.H.)}}}

\address{$^{a}$\:Department of Engineering Mechanics,Northwestern Polytechnical University, Xi’an, 710072, China}

\address{$^{b}$\:School of Engineering, Newcastle University, Newcastle upon Tyne NE1 7RU, UK}

\address{$^{c}$\:Shenzhen Research Institute of Northwestern Polytechnical University, Shenzhen, 518063, China}

\address{$^{d}$\: School of Engineering, University of Liverpool, Liverpool, UK}

\address{$^{e}$\: School of Aeronautics and Astronautics, Sun Yat-sen University, Shenzhen, 518107, Guangdong, China}

\address{$^{f}$\: John A. Paulson School of Engineering and Applied Sciences, Harvard University, Boston, MA, USA}

\date{\today}

\begin{abstract}

Thin elastic flexible structures exploit large geometric deformation to achieve mechanical functionality, making their fracture behavior strongly coupled with the evolving structural configuration. 
This work presents a unified discrete differential geometry (DDG) phase-field framework for fracture in highly deformable thin structures. 
Membrane and bending elasticity are formulated from discrete geometric measures on a triangulated midsurface, while the phase field is defined on the same discrete surface to describe onset of crack growth and subsequent propagation. 
The coupled problem is solved using a staggered scheme, with phase-field irreversibility enforced by an active-set method.
A degradation-deletion procedure removes nearly fully failed elements and reconstructs the DDG topology, enabling complete crack opening and substantial post-fracture reconfiguration.
For two-dimensional in-plane fracture problems, the DDG predictions agree closely with geometrically nonlinear finite element results in both crack evolution and mechanical response.
In three-dimensional tearing, the framework resolves the interaction between out-of-plane deformation and crack propagation, including the experimentally observed convergence and coalescence of initially parallel cracks. 
Its engineering applicability is further demonstrated using an island--bridge structure representative of flexible electronics. The simulations capture direction-dependent deformation modes, instability and snap-through during loading, and the coupled evolution of structural configuration and fracture. 
These results establish the proposed framework as an effective tool for investigating geometry-dependent fracture and supporting the damage-tolerant design of flexible thin structures.

\end{abstract}


\end{frontmatter}

\section{Introduction}

Emerging flexible systems increasingly exploit deformation as a source of functionality rather than merely resisting it~\cite{hebner2026Programmable,bertoldi2017Flexible}.
Thinness and structural compliance allow geometry to become an active design variable: slender interconnects bend and twist to render rigid electronic materials stretchable~\cite{bo2023mechanically,huang2019Assembly,xue2020MechanicallyGuided}, while deliberately programmed buckling, folding, and snapping enable architected structures to change shape, mechanical response, and functionality~\cite{holmesElasticityStabilityShapeshifting2019,azulay2023Instabilityinduced}.
The same principle underlies soft robotic systems, whose ability to adapt, locomote, and interact with complex environments arises from large and continuous changes in configuration~\cite{aziz2026Monolithic}.
Across these applications, flexibility is therefore not merely the capacity to withstand deformation, but the deliberate use of geometric reconfiguration to create mechanical function~\cite{Hysteretic,lu2025Mechanical}.

This functional reliance on geometric freedom, however, also makes failure fundamentally different from fracture in conventional bulk solids.
As a thin structure undergoes large rotations and out-of-plane deformation, its evolving configuration continuously redistributes membrane and bending stresses and thereby alters the driving force and trajectory of a crack~\cite{bayart2010tearing,bayart2011Finitedistance}.
Crack propagation, in turn, releases structural constraints and may trigger further changes in curvature, deformation mode, and load transfer. 
Fracture in flexible thin structures is therefore a coupled process in which structural reconfiguration and crack evolution mutually govern one another~\cite{spada2025coupled,tian2022Mixed,tian2020dynamic}. 
Predicting such behavior consequently requires a fracture framework capable of resolving both evolving crack topology and geometrically nonlinear thin-structure mechanics.

Phase-field fracture provides an effective means of describing complex crack evolution. 
By representing a sharp crack through a continuous scalar field, it naturally accommodates crack initiation, propagation, branching, and coalescence within a fixed computational mesh~\cite{zhuang2022Phase}. 
Its variational foundation was established by Francfort and Marigo~\cite{francfort1998Revisiting} through a reformulation of Griffith's fracture theory and was subsequently regularized by Bourdin et al.~\cite{bourdin2000Numerical}, making the resulting fracture problem amenable to numerical discretization. 
Building on this framework, Amor et al.~\cite{amor2009Regularized} introduced a volumetric-deviatoric decomposition of the elastic energy, while Miehe et al.~\cite{miehe2010Thermodynamically,miehe2010phase} developed a thermodynamically consistent formulation incorporating tension-compression decomposition and crack irreversibility. 

These foundational formulations were largely established for bulk solids under infinitesimal kinematics. 
Their application to flexible thin structures has advanced along two related directions: extending phase-field fracture to finite deformation and coupling it with dimensionally reduced structural models.
Along the first direction, Hesch and Weinberg developed a thermodynamically consistent finite-deformation formulation for general nonlinear materials~\cite{hesch2014Thermodynamically}, 
while Miehe and Schänzel~\cite{miehe2014Phase} coupled finite elasticity with brittle fracture in rubber-like polymers. 
Subsequent studies introduced energy decompositions for nonlinearly elastic solids~\cite{tang2019Phase} and addressed large-strain phenomena such as rate dependence~\cite{loew2019Ratedependent}, elastoplasticity~\cite{dittmann2018Variational}, and near incompressibility~\cite{tian2022Mixed}. 
In parallel, phase-field fracture was extended to plates and shells. 
Following the early plate-shell formulation of Ulmer et al.~\cite{ulmer2012Phase}, Kiendl et al.~\cite{kiendl2016Phasefield} developed phase-field models for Kirchhoff-Love shells using isogeometric analysis.
Proserpio et al.~\cite{proserpio2021Phasefield} combined finite-strain ductile fracture with an isogeometric Kirchhoff-Love formulation. 
More recently, Zhang et al.~\cite{zhang2022Mixed} proposed a mixed Graph–FEM phase-field formulation for nonlinear plates and shells, using a graph Laplacian to discretize fracture evolution on curved surfaces within conventional shell finite elements, while Ma et al.~\cite{ma2024Phasefield} developed a finite-strain ductile fracture model for geometrically exact shells.
Nevertheless, finite-deformation shell fracture requires a consistent representation of large rotations, curvature changes, and crack evolution on an evolving midsurface. Existing formulations often rely on higher-order basis functions, additional degrees of freedom, or through-thickness integration, increasing both formulation complexity and computational cost.

Large rigid-body translations and rotations, strong curvature variations, and substantial post-fracture reconfiguration place demanding requirements on both shell kinematics and spatial discretization~\cite{huang2025tutorial}.
Discrete differential geometry (DDG) provides an alternative approach by constructing thin-shell mechanics directly from geometric quantities defined on triangulated surfaces~\cite{tong2026Discrete}.
Early discrete-shell formulations represented membrane deformation through changes in the surface metric and bending through discrete curvature measures associated with adjacent elements, enabling robust simulation of large rotations and geometrically nonlinear shell deformation~\cite{grinspun2003Discrete,grinspun2006discrete}.
Owing to its local geometric representation and compact discrete degrees of freedom, DDG can also offer high computational efficiency, particularly for simulations involving large deformation and strongly nonlinear structural responses.
Subsequent studies developed variational DDG formulations for the kinematics and mechanics of rods, plates, and shells, in which internal forces and tangent operators are derived consistently from discrete elastic energies~\cite{grinspun2006Computing,clarisse2012discrete}.
These formulations have been further extended to dynamic problems involving buckling, wrinkling, snap-through, contact, and other strongly nonlinear transitions~\cite{huang2025tutorial,tong2026Discrete}.
More recently, DDG-based models have been applied to deployable structures, compliant mechanisms, and soft and origami robotic systems~\cite{zhang2026folding}, where large shape changes, structural elasticity, and actuation dynamics must be resolved within a unified computational framework.

Motivated by these considerations, we develop a unified DDG-based phase-field framework for the fracture simulation in highly deformable thin structures.
By coupling discrete membrane and bending elasticity with phase-field fracture on a common triangulated midsurface, the framework provides a compact formulation for large deformation and crack evolution, while topology updates enable complete separation and post-fracture reconfiguration.
Comparisons with finite element solutions and experimental observations assess its predictive capability.
Applications to flexible island–bridge structures further reveal how structural deformation and fracture mutually influence one another, providing insights into geometry-dependent failure and the damage-tolerant design of flexible systems.

This manuscript is organized as follows. Section 2 formulates the geometrically nonlinear shell-fracture model and presents its Discrete differential geometry and phase-field discretizations. Section 3 describes the numerical solution procedure and topology treatment, while Section 4 evaluates the proposed framework through representative fracture problems involving large deformation and complex structural reconfiguration. Finally, Section 5 summarizes the main conclusions.

\section{Model}
\label{sec:model}

\subsection{Shell geometry and fracture}
\label{sec:continuous-sheet}

\begin{figure}[!htbp]
    \centering
    \includegraphics[
        width=0.9\linewidth,
        trim={0 11cm 0 0},
        clip
    ]{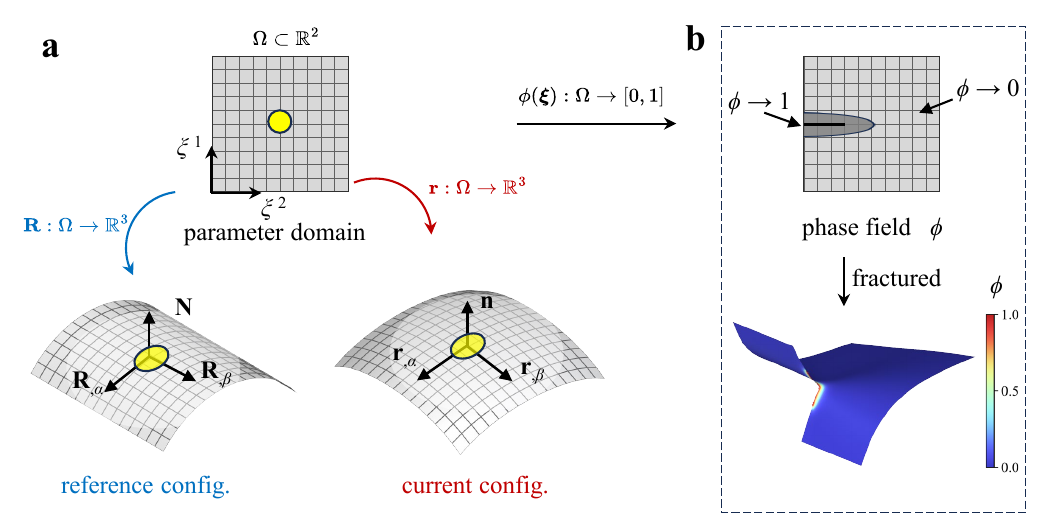}
    \caption {Schematic of the shell kinematics and phase-field fracture formulation: (a) reference and current midsurface configurations mapped from the material parameter domain, and (b) diffuse phase field and the corresponding fractured shell.}
    \label{fig:schematic}
\end{figure}

As illustrated in Fig.~\ref{fig:schematic}(a), let \(\Omega\subset\mathbb{R}^2\) be the material parameter domain of the plate,
with curvilinear coordinates
\(\boldsymbol{\xi}=(\xi^1,\xi^2)\).
The midsurface in the current configuration is described by an embedding
\(\mathbf r:\Omega\rightarrow\mathbb{R}^3\).
At each material point, the surface geometry is characterized by the tangent
vectors \(\mathbf r_{,\alpha}=\partial\mathbf r/\partial\xi^\alpha\)
(\(\alpha=1,2\)) and the unit normal
\(
\mathbf n
=
{ (\mathbf r_{,1}\times\mathbf r_{,2}}) / 
     {\|\mathbf r_{,1}\times\mathbf r_{,2}\|}.
\)
Under the Kirchhoff--Love assumption, a material point through the thickness
is represented as a normal offset from the midsurface,
\begin{equation}
\mathbf s(\xi^1,\xi^2,z)
=
\mathbf r(\xi^1,\xi^2)
+
z\mathbf n(\xi^1,\xi^2),
\qquad
z\in[-h/2,h/2],
\end{equation}
where \(h\) denotes the shell thickness. 

Under the Kirchhoff--Love assumption, we consider the tangential metric on surfaces parallel to the midsurface. 
At a normal offset
\(z\in[-h/2,h/2]\), the differential of the offset surface is
\begin{equation}
\mathrm d\mathbf s
=
\mathrm d\mathbf r+z\,\mathrm d\mathbf n .
\end{equation}  
Its \(2\times2\) metric tensor is therefore
\begin{equation}
\mathbf g(\xi^1,\xi^2,z)
=
(\mathrm d\mathbf r+z\,\mathrm d\mathbf n)^T
(\mathrm d\mathbf r+z\,\mathrm d\mathbf n)
=
\mathbf a(\xi^1,\xi^2)
-2z\mathbf b(\xi^1,\xi^2)
+\mathcal O(z^2),
\label{eq:offset-surface-metric}
\end{equation}  
where terms of second and higher order in the thickness coordinate are neglected\cite{van2017growth}. 
At each material point, the metric tensor encodes the local intrinsic geometry through the lengths and angles of tangent vectors. It can be visualized as the image of an infinitesimal circle in the parameter domain, shown as the yellow metric ellipse in Fig.~\ref{fig:schematic}(a).
The tensors
\begin{equation}
\mathbf a=(\mathrm d\mathbf r)^T\mathrm d\mathbf r,
\qquad
\mathbf b=-(\mathrm d\mathbf r)^T\mathrm d\mathbf n
\label{eq:fundamental form}
\end{equation}  
are the first and second fundamental forms of the midsurface, respectively.
Their components are
\begin{equation}
a_{\alpha\beta}
=
\mathbf r_{,\alpha}\cdot\mathbf r_{,\beta},
\qquad
b_{\alpha\beta}
=
-\mathbf r_{,\alpha}\cdot\mathbf n_{,\beta}
=
\mathbf r_{,\alpha\beta}\cdot\mathbf n,
\qquad \alpha,\beta\in\{1,2\}.
\label{eq:fundamental-form-components}
\end{equation}
The first fundamental form measures lengths and angles on the midsurface,
whereas the second fundamental form measures its curvature.  These two
quadratic forms determine the surface locally.  

As illustrated in Fig.~\ref{fig:schematic}(a), the strain associated with the tangential metric is defined relative to a
stress-free reference configuration. Let
\(\mathbf R(\xi^1,\xi^2)\) denote the reference midsurface, with unit normal
\(\mathbf N(\xi^1,\xi^2)\). 
Its metric is
\begin{equation}
\bar{\mathbf g}(\xi^1,\xi^2,z)
=
\bar{\mathbf a}(\xi^1,\xi^2)
-2z\bar{\mathbf b}(\xi^1,\xi^2)
+\mathcal O(z^2),
\label{eq:reference-metric-gbar}
\end{equation}
where
\[
\bar a_{\alpha\beta}
=
\mathbf R_{,\alpha}\cdot\mathbf R_{,\beta},
\qquad
\bar b_{\alpha\beta}
=
-\mathbf R_{,\alpha}\cdot\mathbf N_{,\beta}.
\]
The deformation of the shell is measured by comparing the tangential metric $\mathbf g$
in the current configuration with that of the stress-free reference
configuration $\bar{\mathbf g}$. 
Geometrically, this comparison is visualized in Fig.~\ref{fig:schematic}(a) by the change in the yellow metric ellipse from the reference to the current configuration.
Using the reference midsurface metric to raise one tensor index,
we define the finite deformation shell strain as
\begin{equation}
    \boldsymbol{\varepsilon}
    =
    \frac{1}{2}\bar{\mathbf g}^{-1}
    \left[
        \mathbf g
        -
        \bar{\mathbf g}
    \right].
    \label{eq:shell-strain}
\end{equation}
Substituting the current and reference metric expansions,
Eqs.~\eqref{eq:offset-surface-metric} and
\eqref{eq:reference-metric-gbar}, into
Eq.~\eqref{eq:shell-strain}, and retaining terms up to first order in the
thickness coordinate \(z\), yields~\cite{van2017growth}
\begin{equation}
    \boldsymbol{\varepsilon}\
    =
    \frac{1}{2}
    \left(
        \bar{\mathbf a}^{-1}\mathbf a
        -\mathbf I
    \right)
    -
    z\bar{\mathbf a}^{-1}
    \left(
        \mathbf b-\bar{\mathbf b}
    \right)
    +\mathcal O(z^2).
    \label{eq:strain-through-thickness}
\end{equation}
Accordingly, the midsurface strain and bending strain are identified as
\begin{equation}
    \boldsymbol{\varepsilon}^{\mathrm m}
    =
    \frac{1}{2}
    \left(
        \bar{\mathbf a}^{-1}\mathbf a
        -\mathbf I
    \right),
    \qquad
    \boldsymbol{\varepsilon}^{\mathrm b}
    =
    \bar{\mathbf a}^{-1}
    \left(
        \mathbf b-\bar{\mathbf b}
    \right),
    \label{eq:membrane-bending-strains}
\end{equation}
To extend the shell formulation to fracture, we introduce a scalar phase field
\(\phi=\phi(\boldsymbol{\xi}):\Omega\rightarrow[0,1]\), defined over the
material coordinates of the reference midsurface
\(\mathbf R(\boldsymbol{\xi})\). The values \(\phi=0\) and \(\phi=1\) represent
intact and fully fractured material, respectively.
As illustrated
in Fig.~\ref{fig:schematic}(b), \(\phi\) remains close to 0 away from the crack,
approaches 1 along the crack, and varies smoothly across a finite-width
diffuse fracture zone. 
Coupling the phase field with the elastic deformation
enables the deformation-driven propagation of cracks to be
resolved.

Let
\(\psi_{\mathrm{el}}^{+}\) and \(\psi_{\mathrm{el}}^{-}\) denote the
crack-driving and crack-insensitive parts of the shell elastic-energy density,
such that
\(\psi_{\mathrm{el}}
=\psi_{\mathrm{el}}^{+}+\psi_{\mathrm{el}}^{-}\).
Only the crack-driving part \(\psi_{\mathrm{el}}^{+}\) is degraded through \cite{zhuang2022Phase,miehe2010phase,miehe2010Thermodynamically}
\begin{equation}
    g(\phi)=(1-\phi)^2+\eta,
    \label{eq:degradation-function}
\end{equation}
where \(0<\eta\ll1\) is a residual-stiffness parameter. The total potential energy of the fractured shell is \cite{amor2009Regularized}
\begin{equation}
    \begin{aligned}
    \Pi[\mathbf r,\phi]
    ={}&
    \int_{\Omega}
    \left[
        g(\phi)\,
        \psi_{\mathrm{el}}^{+}
        (\boldsymbol{\varepsilon})
        +
        \psi_{\mathrm{el}}^{-}
        (\boldsymbol{\varepsilon})
    \right]\mathrm d\bar A
    +
    hG_c
    \int_{\Omega}
    \left[
        \frac{\phi^2}{2\ell}
        +
        \frac{\ell}{2}
        \left|\bar{\nabla}\phi\right|^2
    \right]
    \mathrm d\bar A
    -
    \mathcal W_{\mathrm{ext}}(\mathbf r),
    \end{aligned}
\end{equation}
Here \(G_c\) is the critical fracture energy, \(\ell\) is the phase-field
regularization length, and
\(
    \mathrm d\bar A
    =
    \sqrt{\det\bar{\mathbf a}}\,
    \mathrm d\xi^1\mathrm d\xi^2
\)
is the area element of the reference midsurface, where
\(\bar{\mathbf a}\) denotes its first fundamental form. 
Finally, \(\mathcal W_{\mathrm{ext}}\) denotes the
work of the external loads acting on the current midsurface \(\mathbf r\).
The first integral represents the degraded elastic energy, whereas the second
provides the Griffith fracture energy over the
reference midsurface.

\subsection{Discrete shell formulation}
\label{DDG method}
\begin{figure}[t]
    \centering
    \includegraphics[
        width=0.8\linewidth,
        trim={2cm 11.5cm 0 0},
        clip
    ]{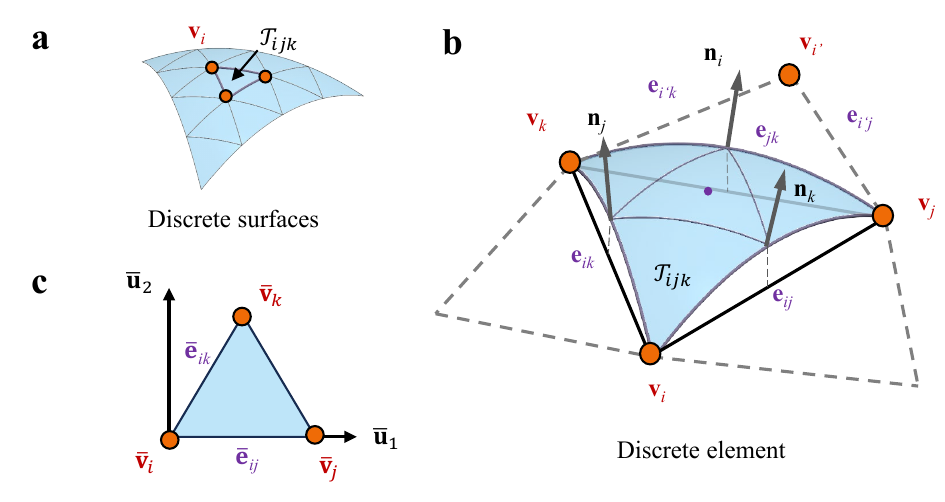}
    \caption {Schematic of the discrete shell geometry. (a) Oriented triangular discretization of the shell midsurface. (b) Discrete shell element consisting of a central triangle and its three edge-adjacent triangles, together with the associated edge vectors and surface normal. (c) Planar representation of the triangle in the local material coordinates.}
    \label{fig:discretized}
\end{figure}

We now construct a discrete counterpart of the continuous shell description
introduced in Section~\ref{sec:continuous-sheet}. 
As illustrated in Fig.~\ref{fig:discretized}(a), the shell midsurface is discretized into an oriented triangular mesh comprising \(N_v\) vertices
\(\mathbf v_i\in\mathbb R^3\) and \(N_t\) triangular elements
\(\mathcal T_{ijk}\). The nodal positions are the only kinematic degrees of
freedom, while rotations, curvatures, and strain measures are derived directly
from the mesh geometry\cite{zhang2026folding,zhang2026Accurate}. Within the DDG framework, these discrete geometric
quantities are used to construct the shell elastic energy. In particular, the
discrete counterparts of the first and second fundamental forms are denoted by
\(\mathbb A\) and \(\mathbb B\), distinguishing them from the continuous tensors
\(\mathbf a\) and \(\mathbf b\).

\subsubsection{Discrete first fundamental form}
As illustrated in Fig.~\ref{fig:discretized}(b), consider a discrete shell element consisting of an oriented triangle
\(\mathcal T_{ijk}=(\mathbf v_i,\mathbf v_j,\mathbf v_k)\).
The two edge vectors emanating from vertex \(i\),
\begin{equation}
    \mathbf e_{ij}=\mathbf v_j-\mathbf v_i,
    \qquad
    \mathbf e_{ik}=\mathbf v_k-\mathbf v_i,
    \label{eq:discrete-edge-basis}
\end{equation}
form a basis of the triangle plane. The discrete first fundamental form is the
metric matrix associated with this basis\cite{clarisse2012discrete,chen2018Physical}:
\begin{equation}
    \mathbb A_{ijk}
    =
    \begin{bmatrix}
        \mathbf e_{ij}\cdot\mathbf e_{ij}
        &
        \mathbf e_{ij}\cdot\mathbf e_{ik}
        \\
        \mathbf e_{ik}\cdot\mathbf e_{ij}
        &
        \mathbf e_{ik}\cdot\mathbf e_{ik}
    \end{bmatrix}.
    \label{eq:discrete-first-form}
\end{equation}
Thus, \(\mathbb A_{ijk}\) is the matrix representation of the triangle metric in the local frame defined by the mesh edges \((\mathbf e_{ij},\mathbf e_{ik})\), providing the discrete
counterpart of the continuous first fundamental form \(\mathbf a\). The area of the triangle can also be derived,
\begin{equation}
    A_e
    =
    \frac{1}{2}\sqrt{\det\mathbb A_{ijk}}.
    \label{eq:discrete-triangle-area}
\end{equation}

\subsubsection{Discrete second fundamental form}
The curvature of a smooth shell is represented by its second fundamental form.
According to Eq.~\eqref{eq:fundamental-form-components}, its components can be
obtained from the variation of the surface normal along tangential directions $\mathrm d\mathbf n$.
On a piecewise-planar triangular mesh, the face normal is constant within each triangle and therefore carries no element-interior curvature. 
The variation of the normal must instead be reconstructed from the relative orientations of adjacent triangles. 
This differs from conventional Kirchhoff--Love shell
finite elements, 
in which bending strains are commonly obtained from second derivatives of the interpolated displacement field or from additional rotational variables.

We adopt the mid-edge construction used in the DDG shell formulation
\citep{grinspun2006Computing,clarisse2012discrete,chen2018Physical}. 
As illustrated in Fig.~\ref{fig:discretized}(b),
for the triangle
\(\mathcal T_{ijk}\), let \(\mathbf n_i\), \(\mathbf n_j\), and
\(\mathbf n_k\) denote the normals associated with the midpoints of the edges
opposite vertices \(i\), \(j\), and \(k\), respectively. 
For an interior edge,
the mid-edge normal is defined as the normalized sum of the consistently
oriented normals of its two incident triangles. 
In particular, consider the edge $\mathbf e_{jk}$ opposite vertex \(i\). For the
interior edge, let \(\mathcal T_{i'kj}\) denote the adjacent triangle, where
\(i'\) is its vertex opposite the shared edge. The corresponding mid-edge unit
normal is defined by
\begin{equation}
    \mathbf n_i
    =
    \frac{
        \mathbf n_{ijk}+\mathbf n_{i'kj}
    }{
        \left\|
        \mathbf n_{ijk}+\mathbf n_{i'kj}
        \right\|
    },
    \label{eq:mid-edge-normal}
\end{equation}
where the consistently oriented, area-weighted face normals are
\begin{equation}
    \mathbf n_{ijk}
    =
    \mathbf e_{ij}\times\mathbf e_{ik},
    \qquad
    \mathbf n_{i'kj}
    =
    \mathbf e_{i'k}\times\mathbf e_{i'j}.
    \label{eq:discrete-face-normal}
\end{equation}
For a boundary edge, we set
\begin{equation}
    \mathbf n_i
    =
    \frac{\mathbf e_{ij}\times\mathbf e_{ik}}{\|\mathbf e_{ij}\times\mathbf e_{ik}\|}.
\end{equation}
The differences between the three mid-edge normals approximate the variation of the normal field over the triangle \cite{clarisse2012discrete}. 
In the edge basis
\((\mathbf e_{ij},\mathbf e_{ik})\), the discrete second fundamental form is
represented by
\begin{equation}
    \mathbb B_{ijk}
    =
    2
    \begin{bmatrix}
        (\mathbf n_j-\mathbf n_i)\cdot\mathbf e_{ij}
        &
        (\mathbf n_j-\mathbf n_i)\cdot\mathbf e_{ik}
        \\[2pt]
        (\mathbf n_k-\mathbf n_i)\cdot\mathbf e_{ij}
        &
        (\mathbf n_k-\mathbf n_i)\cdot\mathbf e_{ik}
    \end{bmatrix}.
    \label{eq:discrete-second-form}
\end{equation}
Let
\(\bar{\mathbf v}_i\), \(\bar{\mathbf v}_j\), and
\(\bar{\mathbf v}_k\) denote the nodal positions of
\(\mathcal T_{ijk}\) in the stress-free reference configuration. Applying
Eqs.~\eqref{eq:discrete-first-form} and
\eqref{eq:discrete-second-form} to these reference nodal positions gives the
reference fundamental forms
\(\bar{\mathbb A}_{ijk}\) and \(\bar{\mathbb B}_{ijk}\), respectively. With
the reference edge vectors
\(\bar{\mathbf e}_{ij}=\bar{\mathbf v}_j-\bar{\mathbf v}_i\) and
\(\bar{\mathbf e}_{ik}=\bar{\mathbf v}_k-\bar{\mathbf v}_i\), the discrete
membrane strain in the edge basis is
\begin{equation}
    \widehat{\mathbb E}^{\mathrm m}_{ijk}
    =
    \frac{1}{2}
    \bar{\mathbb A}_{ijk}^{-1}
    \left(
        \mathbb A_{ijk}-\bar{\mathbb A}_{ijk}
    \right)
    =
    \frac{1}{2}
    \left(
        \bar{\mathbb A}_{ijk}^{-1}\mathbb A_{ijk}
        -
        \mathbb I
    \right),
    \label{eq:discrete-membrane-strain}
\end{equation}
and the discrete bending strain is
\begin{equation}
    \widehat{\mathbb E}^{\mathrm b}_{ijk}
    =
    \bar{\mathbb A}_{ijk}^{-1}
    \left(
        \mathbb B_{ijk}-\bar{\mathbb B}_{ijk}
    \right),
    \label{eq:discrete-bending-strain}
\end{equation}

\subsubsection{Amor split for discrete shell strain}
Let \(\boldsymbol{\varepsilon}\) denote a symmetric two-dimensional strain
tensor expressed in an orthonormal material frame. Introducing
\begin{equation}
    \theta
    =
    \operatorname{tr}(\boldsymbol{\varepsilon}),
    \qquad
    \boldsymbol{\varepsilon}_{\mathrm{dev}}
    =
    \boldsymbol{\varepsilon}
    -
    \frac{1}{2}\theta\mathbf I.
    \label{eq:amor-volumetric-deviatoric-strain}
\end{equation}
the volumetric--deviatoric decomposition of Amor et
al.~\citep{amor2009Regularized} splits the elastic-energy density into
\begin{equation}
    \psi_{\mathrm{el}}^{+}(\boldsymbol{\varepsilon})
    =
    \frac{\lambda+\mu}{2}
    \langle\theta\rangle_{+}^{2}
    +
    \mu\,
    \boldsymbol{\varepsilon}_{\mathrm{dev}}
    :
    \boldsymbol{\varepsilon}_{\mathrm{dev}},
    \qquad
    \psi_{\mathrm{el}}^{-}(\boldsymbol{\varepsilon})
    =
    \frac{\lambda+\mu}{2}
    \langle\theta\rangle_{-}^{2}.
    \label{eq:amor-energy-split}
\end{equation}
Here the positive and negative ramp functions 
$\langle \theta\rangle_{+}=(\theta+|\theta|)/2$ and 
$\langle \theta\rangle_{-}=(\theta-|\theta|)/2$ separate the volumetric
strain into its tensile and compressive contributions. 
The material parameter
\(\mu\) is the shear modulus and
\(\lambda+\mu\) is the two-dimensional bulk modulus. Under plane
stress,
\begin{equation}
    \lambda
    =
    \frac{E\nu}{1-\nu^2},
    \qquad
    \mu
    =
    \frac{E}{2(1+\nu)},
    \label{eq:plane-stress-lame-parameters}
\end{equation}
where \(E\) is the Young's modulus and \(\nu\) is the Poisson ratio.

The standard matrix operations in
Eqs.~\eqref{eq:amor-volumetric-deviatoric-strain} and
\eqref{eq:amor-energy-split} assume that the strain components are expressed
in an orthonormal frame.
The hatted strains in
Eqs.~\eqref{eq:discrete-membrane-strain} and
\eqref{eq:discrete-bending-strain}, however, are matrix representations of
mixed \((1,1)\) strain tensors in the generally non-orthogonal edge basis
\((\bar{\mathbf e}_{ij},\bar{\mathbf e}_{ik})\). 
We therefore construct
\begin{equation}
    \bar{\mathbf u}_1
    =
    \frac{\bar{\mathbf e}_{ij}}
         {\|\bar{\mathbf e}_{ij}\|},
    \qquad
    \bar{\mathbf N}_{ijk}
    =
    \frac{
        \bar{\mathbf e}_{ij}\times\bar{\mathbf e}_{ik}
    }{
        \|\bar{\mathbf e}_{ij}\times\bar{\mathbf e}_{ik}\|
    },
    \qquad
    \bar{\mathbf u}_2
    =
    \bar{\mathbf N}_{ijk}\times\bar{\mathbf u}_1 .
    \label{eq:reference-material-frame}
\end{equation}
on the reference triangle. The matrix
\begin{equation}
    \mathbb M_{ijk}
    =
    \begin{bmatrix}
        \bar{\mathbf e}_{ij}\cdot\bar{\mathbf u}_1
        &
        \bar{\mathbf e}_{ik}\cdot\bar{\mathbf u}_1
        \\
        \bar{\mathbf e}_{ij}\cdot\bar{\mathbf u}_2
        &
        \bar{\mathbf e}_{ik}\cdot\bar{\mathbf u}_2
    \end{bmatrix},
    \label{eq:edge-to-material-map}
\end{equation}
maps vector components from the edge basis to the material frame. Accordingly, the same strain tensors are represented in the
orthonormal material frame by the similarity transformations
\begin{equation}
    \mathbb E^{\mathrm m}_{ijk}
    =
    \mathbb M_{ijk}
    \widehat{\mathbb E}^{\mathrm m}_{ijk}
    \mathbb M_{ijk}^{-1},
    \label{eq:material-membrane-strain}
\end{equation}
\begin{equation}
    \mathbb E^{\mathrm b}_{ijk}
    =
    \mathbb M_{ijk}
    \widehat{\mathbb E}^{\mathrm b}_{ijk}
    \mathbb M_{ijk}^{-1}.
    \label{eq:material-bending-strain}
\end{equation}
Within each triangular element, the membrane and bending strains are assumed to be spatially uniform. Substituting
Eqs.~\eqref{eq:material-membrane-strain} and
\eqref{eq:material-bending-strain} into the elastic energy density and
integrating through the shell thickness gives the element elastic energy
\begin{equation}
    \Pi_{ijk}^{\mathrm{el}}
    =
    \bar A_{e}
    \left\{
        h\left[
            g\left(\phi_{ijk}^{\mathrm{ave}}\right)
            \psi_{\mathrm{el}}^{+}
            \left(\mathbb E_{ijk}^{\mathrm m}\right)
            +
            \psi_{\mathrm{el}}^{-}
            \left(\mathbb E_{ijk}^{\mathrm m}\right)
        \right]
        +
        \frac{h^{3}}{12}
        \psi_{\mathrm{el}}
        \left(\mathbb E_{ijk}^{\mathrm b}\right)
    \right\},
    \label{eq:discrete-decomposed-elastic-energy}
\end{equation}
where
$\phi_{ijk}^{\mathrm{ave}}=(\phi_i+\phi_j+\phi_k)/3$ 
is the element-averaged phase-field value, and $\bar A_{e}$ is area of the triangle in reference configuration.
For the flexible thin structures considered here, fracture is driven primarily
by the tensile membrane energy, while the compressive membrane and bending
energies remain undegraded. 
As an additional numerical treatment, these
contributions are locally degraded in nearly fully damaged elements to eliminate
residual traction and allow crack opening.

\subsection{Phase-field discretization}
\label{sec:phase-field-discretization}

The phase field is discretized at the vertices of the same triangular mesh used for the shell geometry. For a reference triangle
\(\mathcal T_{ijk}\), the corresponding nodal phase-field vector is defined as
\begin{equation}
    \boldsymbol{\phi}_{ijk}
    =
    \begin{bmatrix}
        \phi_i & \phi_j & \phi_k
    \end{bmatrix}^{T}.
    \label{eq:element-phase-field-vector}
\end{equation}
For each reference triangle \(\mathcal T_{ijk}\), we introduce local Cartesian material coordinates \((\bar x,\bar y)\) in the orthonormal frame
\((\bar{\mathbf u}_1,\bar{\mathbf u}_2)\) defined in
Eq.~\eqref{eq:reference-material-frame}. The coordinates of vertex
\(a\in\{i,j,k\}\), measured relative to vertex \(i\), are defined as
\begin{equation}
    \bar x_a
    =
    \bar{\mathbf u}_1\cdot
    (\bar{\mathbf v}_a-\bar{\mathbf v}_i),
    \qquad
    \bar y_a
    =
    \bar{\mathbf u}_2\cdot
    (\bar{\mathbf v}_a-\bar{\mathbf v}_i).
    \label{eq:local-material-coordinates}
\end{equation}
The phase field is linearly interpolated over the triangle as
\begin{equation}
    \phi
    =
    \mathbb N_{ijk}^{\phi}
    \boldsymbol{\phi}_{ijk},
    \qquad
    \mathbb N_{ijk}^{\phi}
    =
    \begin{bmatrix}
        N_i & N_j & N_k
    \end{bmatrix},
    \label{eq:linear-phase-field-interpolation}
\end{equation}
where the linear shape functions at an arbitrary point
\((\bar x,\bar y)\in\mathcal T_{ijk}\) are
\begin{align}
    N_i(\bar x,\bar y)
    &=
    1+
    \frac{
        (\bar y_j-\bar y_k)\bar x
        +
        (\bar x_k-\bar x_j)\bar y
    }{2\bar A_{ijk}},
    \\
    N_j(\bar x,\bar y)
    &=
    \frac{
        \bar y_k\bar x-\bar x_k\bar y
    }{2\bar A_{ijk}},
    \qquad
    N_k(\bar x,\bar y)
    =
    \frac{
        -\bar y_j\bar x+\bar x_j\bar y
    }{2\bar A_{ijk}}.
    \label{eq:linear-phase-field-shape-functions}
\end{align}
Differentiating the shape functions with respect to the local material
coordinates gives the constant element gradient operator
\begin{equation}
    \bar{\mathbb D}_{ijk}^{\phi}
    =
    \begin{bmatrix}
        \displaystyle\frac{\partial N_i}{\partial\bar x}
        &
        \displaystyle\frac{\partial N_j}{\partial\bar x}
        &
        \displaystyle\frac{\partial N_k}{\partial\bar x}
        \\[2mm]
        \displaystyle\frac{\partial N_i}{\partial\bar y}
        &
        \displaystyle\frac{\partial N_j}{\partial\bar y}
        &
        \displaystyle\frac{\partial N_k}{\partial\bar y}
    \end{bmatrix}
    =
    \frac{1}{2\bar A_{ijk}}
    \begin{bmatrix}
        \bar y_j-\bar y_k
        &
        \bar y_k
        &
        -\bar y_j
        \\
        \bar x_k-\bar x_j
        &
        -\bar x_k
        &
        \bar x_j
    \end{bmatrix}.
    \label{eq:element-phase-field-gradient-operator}
\end{equation}
Accordingly, the phase-field gradient within the triangle  
\(\mathcal T_{ijk}\) is
\begin{equation}
    \bar{\nabla}\phi
    =
    \bar{\mathbb D}_{ijk}^{\phi}
    \boldsymbol{\phi}_{ijk},
    \qquad
    \left|\bar{\nabla}\phi\right|^2
    =
    \boldsymbol{\phi}_{ijk}^{T}
    \left(\bar{\mathbb D}_{ijk}^{\phi}\right)^{T}
    \bar{\mathbb D}_{ijk}^{\phi}
    \boldsymbol{\phi}_{ijk}.
    \label{eq:discrete-phase-field-gradient}
\end{equation}
For the AT2 phase-field model, the regularized fracture energy associated
with triangle \(\mathcal T_{ijk}\) is expressed as
\begin{equation}
    \Pi_{ijk}^{\mathrm{frac}}
    =
    hG_c
    \int_{\mathcal T_{ijk}}
    \left[
        \frac{\phi^2}{2\ell}
        +
        \frac{\ell}{2}
        \left|\bar{\nabla}\phi\right|^2
    \right]
    \mathrm d\bar A,
    \label{eq:element-fracture-energy-integral}
\end{equation}
Using the interpolation relation
in Eq.~\eqref{eq:linear-phase-field-interpolation}, the local crack-density
term satisfies
\begin{equation}
    \phi^2
    =
    \boldsymbol{\phi}_{ijk}^{T}
    \left(\mathbb N_{ijk}^{\phi}\right)^{T}
    \mathbb N_{ijk}^{\phi}
    \boldsymbol{\phi}_{ijk}.
    \label{eq:discrete-local-crack-density}
\end{equation}
Substituting Eqs.~\eqref{eq:discrete-local-crack-density} and
\eqref{eq:discrete-phase-field-gradient} into
Eq.~\eqref{eq:element-fracture-energy-integral} yields
\begin{equation}
    \Pi_{ijk}^{\mathrm{frac}}
    =
    \frac{hG_c}{2\ell}
    \boldsymbol{\phi}_{ijk}^{T}
    \mathbb M_{ijk}^{\phi}
    \boldsymbol{\phi}_{ijk}
    +
    \frac{hG_c\ell}{2}
    \boldsymbol{\phi}_{ijk}^{T}
    \mathbb K_{ijk}^{\phi}
    \boldsymbol{\phi}_{ijk},
    \label{eq:discrete-element-fracture-energy}
\end{equation}
where 
\begin{align}
    \mathbb M_{ijk}^{\phi}
    &=
    \int_{\mathcal T_{ijk}}
    \left(\mathbb N_{ijk}^{\phi}\right)^{T}
    \mathbb N_{ijk}^{\phi}
    \,\mathrm d\bar A,
    \label{eq:element-phase-field-mass-matrix}
    \\
    \mathbb K_{ijk}^{\phi}
    &=
    \int_{\mathcal T_{ijk}}
    \left(\bar{\mathbb D}_{ijk}^{\phi}\right)^{T}
    \bar{\mathbb D}_{ijk}^{\phi}
    \,\mathrm d\bar A.
    \label{eq:element-phase-field-gradient-matrix}
\end{align}
The matrices \(\mathbb M_{ijk}^{\phi}\) and
\(\mathbb K_{ijk}^{\phi}\) represent the discrete local crack-density
and gradient-regularization contributions, respectively.
For a linear triangular element, exact integration gives
\begin{equation}
    \mathbb M_{ijk}^{\phi}
    =
    \frac{\bar A_{e}}{12}
    \begin{bmatrix}
        2 & 1 & 1 \\
        1 & 2 & 1 \\
        1 & 1 & 2
    \end{bmatrix}.
    \label{eq:explicit-element-phase-field-mass-matrix}
\end{equation}
where $\bar A_{e} $ is the area of the reference configuration.
Moreover, because
\(\bar{\mathbb D}_{ijk}^{\phi}\) is constant within the triangle,
the gradient matrix reduces to
\begin{equation}
    \mathbb K_{ijk}^{\phi}
    =
    \bar A_{e}
    \left(\bar{\mathbb D}_{ijk}^{\phi}\right)^{T}
    \bar{\mathbb D}_{ijk}^{\phi}.
    \label{eq:explicit-element-phase-field-gradient-matrix}
\end{equation}

\section{Numerical implementation}
\label{sec:numerical-implementation}

\subsection{Discrete variables and staggered formulation}
\label{sec:discrete-staggered-formulation}

Let \(N_v\) be the current number of mesh vertices.  The mechanical unknowns
are the three-dimensional coordinates of all vertices,
\begin{equation}
    \mathbf x
    =
    \begin{bmatrix}
        \mathbf v_1 &
        \mathbf v_2 &
        \cdots &
        \mathbf v_{N_v}
    \end{bmatrix}
    \in\mathbb R^{3N_v},
    \qquad
    \mathbf v_a=
    \begin{bmatrix}x_a&y_a&z_a\end{bmatrix},
    \label{eq:global-displacement-dofs}
\end{equation}
whereas the nodal phase-field unknowns are collected in
\begin{equation}
    \boldsymbol{\phi}
    =
    \begin{bmatrix}
        \phi_1&\phi_2&\cdots&\phi_{N_v}
    \end{bmatrix}
    \in\mathbb R^{N_v}.
    \label{eq:global-phase-dofs}
\end{equation}
The complete discrete state therefore contains \(4N_v\) scalar degrees of
freedom,
\begin{equation}
    \mathbf q
    =
    \begin{bmatrix}
        \mathbf x&\boldsymbol{\phi}
    \end{bmatrix}^{T}.
    \label{eq:complete-discrete-state}
\end{equation}
Although the mechanical coordinates and the phase field jointly define the
discrete state, their governing equations are not solved as a monolithic
system. For improved nonlinear robustness, the coupled problem is partitioned
into a mechanical subproblem and a phase-field subproblem, which are advanced
sequentially at each time step \cite{amor2009Regularized,seles2019residual,zhuang2022Phase}.

\subsubsection{Mechanical subproblem}

At time $t_{n+1}$, the mechanical configuration is determined with
the phase field fixed at its previously accepted value
$\boldsymbol{\phi}_n$. The corresponding discrete mechanical potential
is defined as
\begin{equation}
    \Pi^{\mathrm{el}}
    \left(
        \mathbf x;
        \boldsymbol{\phi}_n
    \right)
    =
    \sum_{\mathcal T_{ijk}}^{N_t}
    \Pi_{ijk}^{\mathrm{el}}
    \left(
        \mathbf x;
        \boldsymbol{\phi}_n
    \right)
    -
    \mathcal W_{\mathrm{ext},n+1}(\mathbf x),
    \label{eq:discrete-mechanical-potential}
\end{equation}
where the summation is taken over the $N_t$ oriented triangles of the
current mesh. Accounting for inertia and mechanical damping, the
configuration $\mathbf x_{n+1}$ is obtained from
\begin{equation}
    \mathbf R^{\mathrm{el}}
    \left(
        \mathbf x_{n+1};
        \boldsymbol{\phi}_n
    \right)
    :=
    \nabla_{\mathbf x}
    \Pi^{\mathrm{el}}
    \left(
        \mathbf x_{n+1};
        \boldsymbol{\phi}_n
    \right)
    +
    \mathbb M^{x}\ddot{\mathbf x}_{n+1}
    +
    \mathbb C^{x}\dot{\mathbf x}_{n+1}
    =
    \mathbf 0,
    \label{eq:discrete-mechanical-equilibrium}
\end{equation}
where $\mathbb M^{x}$ and $\mathbb C^{x}$ denote the mechanical mass
and damping matrices, respectively.
Equation~\eqref{eq:discrete-mechanical-equilibrium} is solved by the
Newton--Raphson method. Let \(\mathbf x_{n+1}^{(m)}\) denote the mechanical
configuration at the \(m\)-th iteration. The residual vector and the
consistent tangent stiffness matrix are evaluated as
\begin{equation}
    \mathbf R^{\mathrm{el}}_{(m)}
    =
    \mathbf R^{\mathrm{el}}
    \left(
        \mathbf x_{n+1}^{(m)};
        \boldsymbol{\phi}_n
    \right),
    \qquad
    \mathbf K^{\mathrm{el}}_{(m)}
    =
    \frac{\partial\mathbf R^{\mathrm{el}}}
         {\partial\mathbf x}
    .
    \label{eq:mechanical-residual-tangent}
\end{equation}
The mechanical correction is obtained from
\begin{equation}
    \mathbf K^{\mathrm{el}}_{(m)}
    \Delta\mathbf x^{(m)}
    =
    \mathbf R^{\mathrm{el}}_{(m)},
    \qquad
    \mathbf x_{n+1}^{(m+1)}
    =
    \mathbf x_{n+1}^{(m)}
    -
    \Delta\mathbf x^{(m)},
    \label{eq:mechanical-newton-update}
\end{equation}
and the iterations are continued until the prescribed convergence criteria
are satisfied.

\subsubsection{Phase-field subproblem}
After convergence of the mechanical subproblem, the tensile membrane energy
is transferred to an element-wise history variable\cite{miehe2010Thermodynamically}. For each triangle
\(\mathcal T_{ijk}\), the update is defined by
\begin{equation}
    H_{ijk}^{n+1}
    =
    \max
    \left\{
        H_{ijk}^{n},
        h\,
        \psi_{\mathrm{el}}^{+}
        \left(
            \mathbb E_{ijk}^{\mathrm m}
            \left(\mathbf x_{n+1}\right)
        \right)
    \right\},
    \label{eq:element-history-update}
\end{equation}
These history variables are initialized to zero at the beginning of the computation. Accordingly, fracture within each element is driven by the tensile deformation of the midsurface.
With the converged configuration $\mathbf x_{n+1}$ and the updated
history field held fixed, the viscously regularized phase-field
potential is defined as
\begin{equation}
    \begin{aligned}
    \Pi^{\mathrm{frac}}
    \left(
        \boldsymbol{\phi}_{n+1};
        \mathbf H_{n+1}
    \right)
    ={}&
    \sum_{\mathcal T_{ijk}}^{N_t}
    \left[
        \bar A_e\,
        g(\boldsymbol{\phi}_{n+1})
        H_{ijk}^{n+1}
        +
        \Pi_{ijk}^{\mathrm{frac}}
        \left(
            \boldsymbol{\phi}_{n+1}
        \right)
    \right]
    \\
    &+
    \frac{\eta_{\phi}}{2\Delta t}
    \left(
        \boldsymbol{\phi}_{n+1}
        -
        \boldsymbol{\phi}_n
    \right)^T
    \mathbb M^{\phi}
    \left(
        \boldsymbol{\phi}_{n+1}
        -
        \boldsymbol{\phi}_n
    \right),
    \end{aligned}
    \label{eq:discrete-phase-potential}
\end{equation}
where $\mathbf H_{n+1}$ is the global vector of $H_{ijk}^{n+1}$,
$\eta_{\phi}$ is the phase-field viscosity,which is set to 0.01 throughout this work. and
$\mathbb M^{\phi}$ is the global phase-field mass matrix.
The corresponding phase-field residual is defined as
\begin{equation}
    \mathbf R^{\mathrm{frac}}
    \left(
        \boldsymbol{\phi}_{n+1};
        \mathbf H_{n+1}
    \right)
    :=
    \nabla_{\boldsymbol{\phi}}
    \Pi^{\mathrm{frac}}
    \left(
        \boldsymbol{\phi}_{n+1};
        \mathbf H_{n+1}
    \right).
    \label{eq:discrete-phase-residual}
\end{equation}
Let \(\boldsymbol{\phi}_{n+1}^{(m)}\) denote the phase-field solution at the \(m\)-th iteration. The residual vector and tangent matrix are evaluated as
\begin{equation}
    \mathbf R^{\mathrm{frac}}_{(m)}
    =
    \mathbf R^{\mathrm{frac}}
    \left(
        \boldsymbol{\phi}_{n+1}^{(m)};
        \mathbf H_{n+1}
    \right),
    \qquad
    \mathbf K^{\mathrm{frac}}_{(m)}
    =
    \frac{\partial\mathbf R^{\mathrm{frac}}}
         {\partial\boldsymbol{\phi}}
    \left(
        \boldsymbol{\phi}_{n+1}^{(m)};
        \mathbf H_{n+1}
    \right).
    \label{eq:phase-residual-tangent}
\end{equation}
The phase-field correction is obtained from
\begin{equation}
    \mathbf K^{\mathrm{frac}}_{(m)}
    \Delta\boldsymbol{\phi}^{(m)}
    =
    \mathbf R^{\mathrm{frac}}_{(m)},
    \qquad
    \boldsymbol{\phi}_{n+1}^{(m+1)}
    =
    \boldsymbol{\phi}_{n+1}^{(m)}
    -
    \Delta\boldsymbol{\phi}^{(m)}.
    \label{eq:phase-field-update}
\end{equation}
Fracture evolution imposes two additional constraints on the phase-field
variable: irreversibility and boundedness. Accordingly, each unconstrained
nodal degree of freedom $a$ must satisfy
\begin{equation}
    0
    \leq
    \phi_{a,n}
    \leq
    \phi_a
    \leq
    1,
    \label{eq:phase-field-bounds}
\end{equation}
The lower bound prevents crack healing, whereas the upper bound
restricts the phase field to its physically admissible range. Consequently,
the update \(\boldsymbol{\phi}_{n+1}\) is obtained from a bound-constrained
phase-field problem, which is solved using the active-set method.

\subsection{Active-set method}
\label{sec:active-set}

The bound constraints are enforced using an active-set method\cite{heister2015primaldual, loew2019Ratedependent, tian2020dynamic}.
Since $\mathbf R^{\mathrm{frac}}$ is the gradient of the incremental
phase-field functional, a trial change along its negative direction
may reduce the functional but may also violate the admissible bounds
$[\phi_{a,n},1]$.
To incorporate these bounds, we define the projected residual at each
node $a$ as
\begin{equation}
    \widehat R_a
    =
    \phi_a -
    \min\left\{
        1,\max\left\{\phi_{a,n},z_a\right\}
    \right\},
    \qquad
    z_a = \phi_a - c_a R_a^{\mathrm{frac}},
    \qquad
    c_a > 0.
    \label{eq:projected-residual}
\end{equation}
where $z_a$ denotes the trial phase-field value  and $c_a$ is a positive
scaling factor. Here, $\phi_a$ denotes the current nodal phase-field
value.
The projection clips the trial value to the admissible interval.
Depending on whether $z_a$ lies below, within, or above this interval,
the projected residual becomes
\begin{equation}
    \widehat R_a
    =
    \begin{cases}
        \phi_a-\phi_{a,n},
        & z_a\leq\phi_{a,n},\\[2pt]
        c_aR_a^{\mathrm{frac}},
        & \phi_{a,n}<z_a<1,\\[2pt]
        \phi_a-1,
        & z_a\geq1.
    \end{cases}
    \label{eq:projected-residual-branches}
\end{equation}
These three branches define the lower
active set $\mathcal A_-$, the free set $\mathcal F$, and the upper
active set $\mathcal A_+$, respectively.
Setting $\widehat R_a=0$ therefore enforces the lower bound
$\phi_a=\phi_{a,n}$ on $\mathcal A_-$, the physical equilibrium
condition $R_a^{\mathrm{frac}}=0$ on $\mathcal F$, and the upper bound
$\phi_a=1$ on $\mathcal A_+$.
Thus, the equation $\widehat{\mathbf R}=\mathbf 0$ combines the
active bound constraints and the free-node equilibrium equations
in a single system.
the resulting projected residual admits a semismooth Newton
interpretation.

In the numerical implementation, the trial phase-field values are
evaluated before each Newton step, and the nodal degrees of freedom
are partitioned into the lower active set $\mathcal A_-$, the free
set $\mathcal F$, and the upper active set $\mathcal A_+$.
The projected residual
$\widehat{\mathbf R}$ then replaces the original residual in the
Newton system of Eq.~\eqref{eq:phase-residual-tangent}.
For nodes in either active
set, the projected residual reduces to the associated bound equation.
Accordingly, the corresponding tangent row is replaced by a unit row,
and the right-hand side is set to the correction required to reach
that bound. 
For nodes in the free set, the projected residual and its tangent
reduce to their original, unprojected forms.

The scaling factor $c_a$ controls the trial phase-field value $z_a$
and therefore influences the identification of the active sets and
the convergence behavior of the iteration. To account for differences
in the local tangent stiffness, the physical residual is normalized
by the magnitude of the corresponding diagonal tangent entry. In the
present implementation,
\begin{equation}
    c_a
    =
    \frac{\alpha_H}
    {\left|K_{aa}^{\mathrm{frac}}\right|+\varepsilon_D},
    \qquad
    \alpha_H=0.1,
    \label{eq:projected-residual-scaling}
\end{equation}
where $\varepsilon_D$ is a small positive regularization parameter
that prevents division by zero. After each Newton solve, the phase
field is updated and the active sets are reassessed, allowing nodes
to enter or leave either active set. Convergence is monitored using
the norm of $\widehat{\mathbf R}$, which accounts simultaneously for
the physical equilibrium of the free nodes and satisfaction of the
active bound constraints.

\subsection{Degradation and deletion of failed elements}
\label{sec:element-deletion}

The phase-field formulation replaces a sharp crack by a diffuse damage zone.
However, once fracture is fully developed, the damaged region should rapidly
lose its mechanical response and the diffuse crack representation should be
converted into an explicit topological separation. 
This transition is handled
through a degradation--deletion treatment of nearly failed elements.

Before deletion, the residual stiffness associated with the undegraded
compressive contribution is further suppressed in elements approaching
complete failure. 
The tension--compression split is introduced to prevent
compressive loading from driving crack growth in the small-deformation
regime. 
Once a crack has formed, the surrounding elements effectively become part of the newly created boundary, so retaining their residual compressive contribution is no longer necessary. Degrading this contribution facilitates a stable transition to element deletion.

After convergence of the staggered solution at \(t_{n+1}\), an oriented
triangle \(\mathcal T_{ijk}\) is classified as failed when
\begin{equation}
    \phi_{ijk}^{\mathrm{ave}}
    =
    \frac{\phi_i+\phi_j+\phi_k}{3}
    \geq
    \phi_{\mathrm{del}} .
    \label{eq:element-deletion-criterion}
\end{equation}
Here, \(\phi_{\mathrm{del}}\) is the threshold for identifying failed elements and triggering their deletion, and is typically chosen in the range 0.95–0.99.

Following deletion, unused vertices are discarded and the remaining vertices
and triangles are renumbered. The reference and current coordinates,
velocities, phase-field values and increments, prescribed constraints, and
element-wise history variables are transferred through the corresponding
old-to-new index maps. Since the DDG membrane and bending operators rely on
consistent triangle orientation and edge-based face adjacency, the mesh
connectivity is maintained and reconstructed using a half-edge data
structure\cite{crane2018discrete}.

\subsection{Solution scheme}
\label{sec:solution-scheme}

The computational meshes are generated using COMSOL Multiphysics 6.1 \cite{comsol61} and subsequently imported into an in-house solver implemented in C++.
The resulting sparse linear systems are solved using the parallel PARDISO sparse direct solver \cite{schenk2004pardiso} provided by Intel oneAPI Math Kernel Library \cite{intelOneMKL}.
For completeness, the overall numerical procedure is summarized in
Algorithm~\ref{alg:staggered-solution}.
At each time step, the mechanical and phase-field subproblems are
solved sequentially. The converged mechanical solution is used to
update the history field, after which the phase-field problem is
solved subject to the irreversibility and boundedness constraints.
Finally, failed elements are removed and the associated mesh topology
is reconstructed before advancing to the next time step.
Although Algorithm~\ref{alg:staggered-solution} presents a single
mechanical--phase-field staggered cycle for clarity, multiple
staggered cycles may be performed within a time step when rapid
fracture evolution requires stronger coupling between the two
subproblems.
The source code of the simulation framework is publicly available in an open-access repository \cite{Zhang2026_PhaseField}.

\begin{algorithm}[t]
\caption{Staggered solution procedure}
\label{alg:staggered-solution}
\begin{algorithmic}[1]
\Require Initial mesh, $\mathbf{x}_0$, $\boldsymbol{\phi}_0$,
         $\mathbf{H}_0$
\For{$n=0,\ldots,N-1$}
    \State Apply the loading at $t_{n+1}$
    \State Predict the mechanical configuration $\mathbf{x}_{n+1}$

    \Statex
    \State \textit{Mechanical subproblem}
    \While{the mechanical field has not converged}
        \State Assemble the mechanical residual and tangent matrix
        \State Solve the mechanical Newton system
        \State Update $\mathbf{x}_{n+1}$
    \EndWhile
    
    \Statex
    \State Update the history field $\mathbf{H}_{n+1}$

    \Statex
    \State \textit{Phase-field subproblem}
    \While{the phase field has not converged}
        \State Assemble the phase-field residual and tangent matrix
        \State Construct the projected residual and identify the active sets
        \State Solve the active-set Newton system
        \State Update $\boldsymbol{\phi}_{n+1}$
    \EndWhile
    
    \Statex
    \State Update failed elements and reconstruct the mesh topology
\EndFor
\Ensure $\mathbf{x}_N$, $\boldsymbol{\phi}_N$, and the final mesh
\end{algorithmic}
\end{algorithm}

\section{Numerical results}

This section presents four numerical examples to systematically assess the proposed DDG–phase-field computational framework, progressing from fundamental verification to a representative engineering application.
First, a tensile test on a pre-cracked plate is compared with reference finite element solutions to verify the numerical implementation. 
A perforated cracked plate is then examined to demonstrate the necessity of a nonlinear elastic description in capturing the interaction between finite deformation, stress concentration, and crack propagation. 
Subsequently, a three-dimensional tearing test demonstrates the coupling of the geometrically exact DDG formulation with phase-field fracture under finite deformation. 
Finally, the framework is applied to an island–bridge structure in flexible electronics to investigate the effects of loading direction, loading magnitude, and pre-existing cracks on its fracture behavior.

\subsection{Tensile test of a pre-cracked plate}

\begin{figure}[h]
    \centering
    \includegraphics[
        width=0.9\linewidth,
        trim={0 6cm 0 0},
        clip
    ]{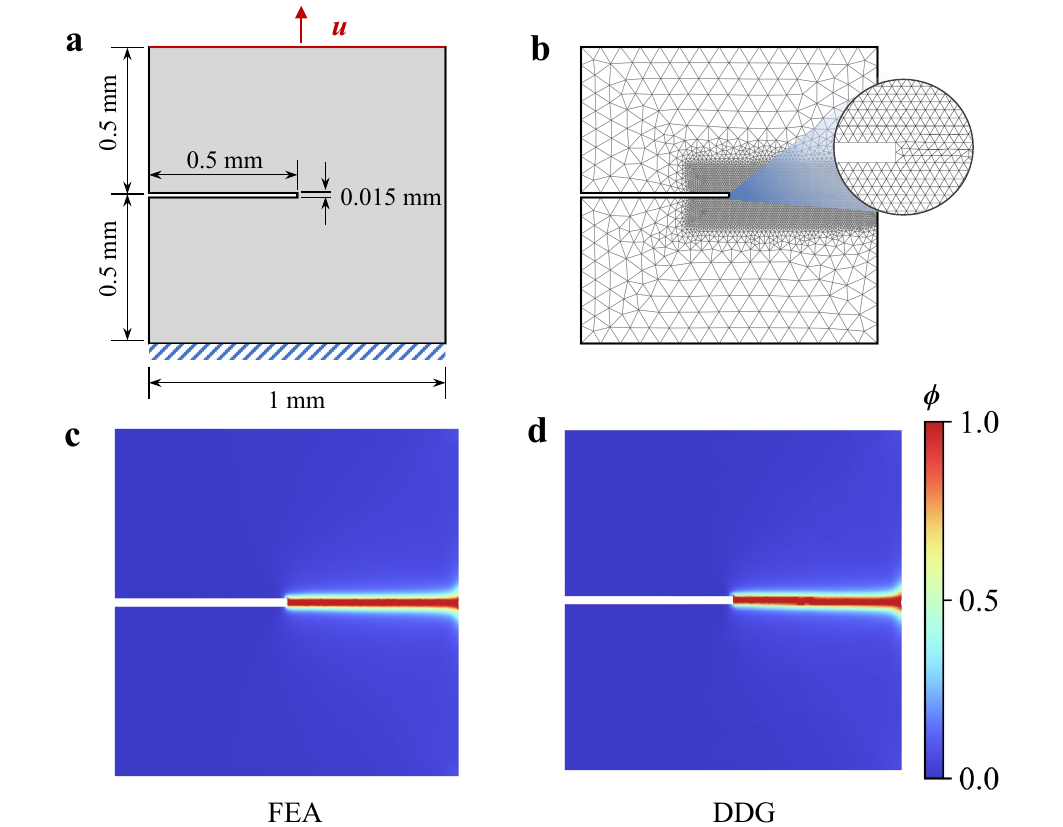}
    \caption{Tensile fracture of a pre-cracked  square plate.
    (a) Geometry and boundary conditions.
    (b) Triangular mesh with local refinement along the anticipated crack path and an enlarged view of the refined region.
    Phase-field distributions predicted by (c) finite element analysis (FEA) and (d) the discrete differential geometry (DDG) method.
    }
    \label{fig:crackPlate}
\end{figure}

A two-dimensional single-edge-notched square plate subjected to tension is first considered to verify the proposed DDG--phase-field framework.
The geometry and boundary conditions are identical to those of the classical benchmark presented in~\cite{hesch2014Thermodynamically,miehe2010Thermodynamically}, while the reference FEA solution is obtained using the phase-field fracture functionality in COMSOL Multiphysics~6.1.
As illustrated in Fig.~\ref{fig:crackPlate}, the plate contains a horizontal pre-crack extending from the midpoint of its left boundary and is loaded by a prescribed vertical displacement.
The material parameters are Young's modulus $E=210~\mathrm{GPa}$, Poisson's ratio $\nu=0.3$, and critical energy release rate $G_c=2700~\mathrm{J/m^2}$~\cite{zhuang2022Phase}.
The computational domain is discretized using triangular elements and locally refined along the anticipated crack-propagation region.
The phase-field length scale is set to $\ell_0=1.5\times10^{-2}~\mathrm{mm}$, and the maximum element size in the refined region is $h=7.5\times10^{-3}~\mathrm{mm}$.
To ensure a consistent comparison, the DDG and FEA models employ identical fracture parameters, mesh resolution, and loading increments.
As shown in Fig.~\ref{fig:crackPlate}(c) and (d). the phase-field distributions predicted by DDG and FEA agree closely.

\begin{figure}[t]
    \centering
    \includegraphics[
        width=0.9\linewidth,
        trim={0 13.5cm 0 0},
        clip
    ]{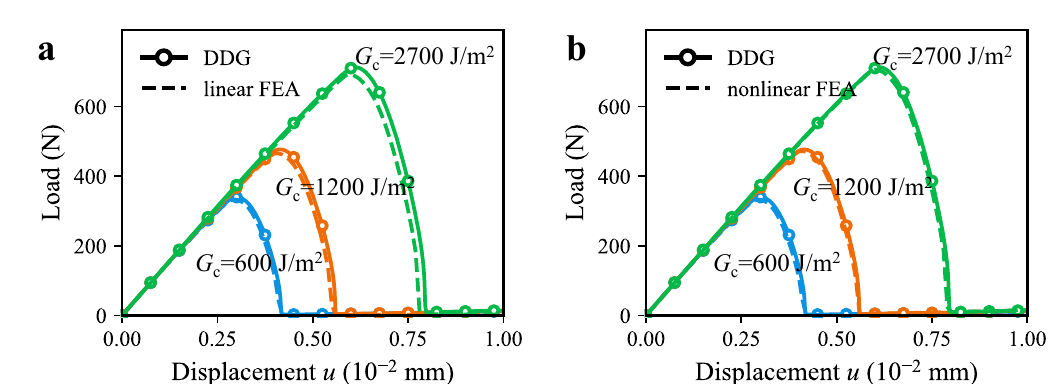}
    \caption{Load--displacement curves of the pre-cracked plate for different $G_c$, comparing DDG with (a) linear and (b) geometrically nonlinear FEA.}
    \label{fig:crackPlate_Loadcurve}
\end{figure}

The agreement between the DDG and FEA solutions is not limited to the predicted crack pattern but is also evident in the quantitative mechanical response.
Fig.~\ref{fig:crackPlate_Loadcurve} compares the load--displacement responses predicted by DDG with those obtained using linear and geometrically nonlinear FEA for different values of $G_c$.
For $G_c=2700~\mathrm{J/m^2}$, the load initially increases almost linearly with the prescribed displacement $u$ and reaches its maximum shortly before fracture occurs at approximately $u=6\times10^{-3}~\mathrm{mm}$.
The subsequent crack propagation produces a rapid loss of load-carrying capacity, with the reaction force decreasing to nearly zero.
Reducing $G_c$ promotes earlier crack growth and consequently shifts both the peak load and the post-peak response towards smaller displacements.

As shown in Fig.~\ref{fig:crackPlate_Loadcurve}(a), for $G_c=600~\mathrm{J/m^2}$, fracture occurs at a relatively small deformation, and the load--displacement responses predicted by linear FEA and DDG remain in close agreement.
As $G_c$ increases, fracture is delayed and larger deformations develop before crack propagation, leading to a progressively more noticeable discrepancy between the two solutions.
This deformation-dependent deviation is primarily associated with the treatment of geometric nonlinearity: the DDG formulation inherently accounts for the evolving geometry, whereas the linear FEA model is based on a small-deformation approximation.
To isolate this effect, Fig.~\ref{fig:crackPlate_Loadcurve}(b) compares the DDG results with those obtained from geometrically nonlinear FEA.
The discrepancy observed in the linear analysis is then largely eliminated, and close agreement is recovered over the entire load--displacement response for all three values of $G_c$.

\subsection{Fracture of a perforated plate under finite deformation}
\label{sec:perforated-plate}

\begin{figure}[t]
    \centering
    \includegraphics[
        width=0.9\linewidth,
        trim={0 3cm 0 0},
        clip
    ]{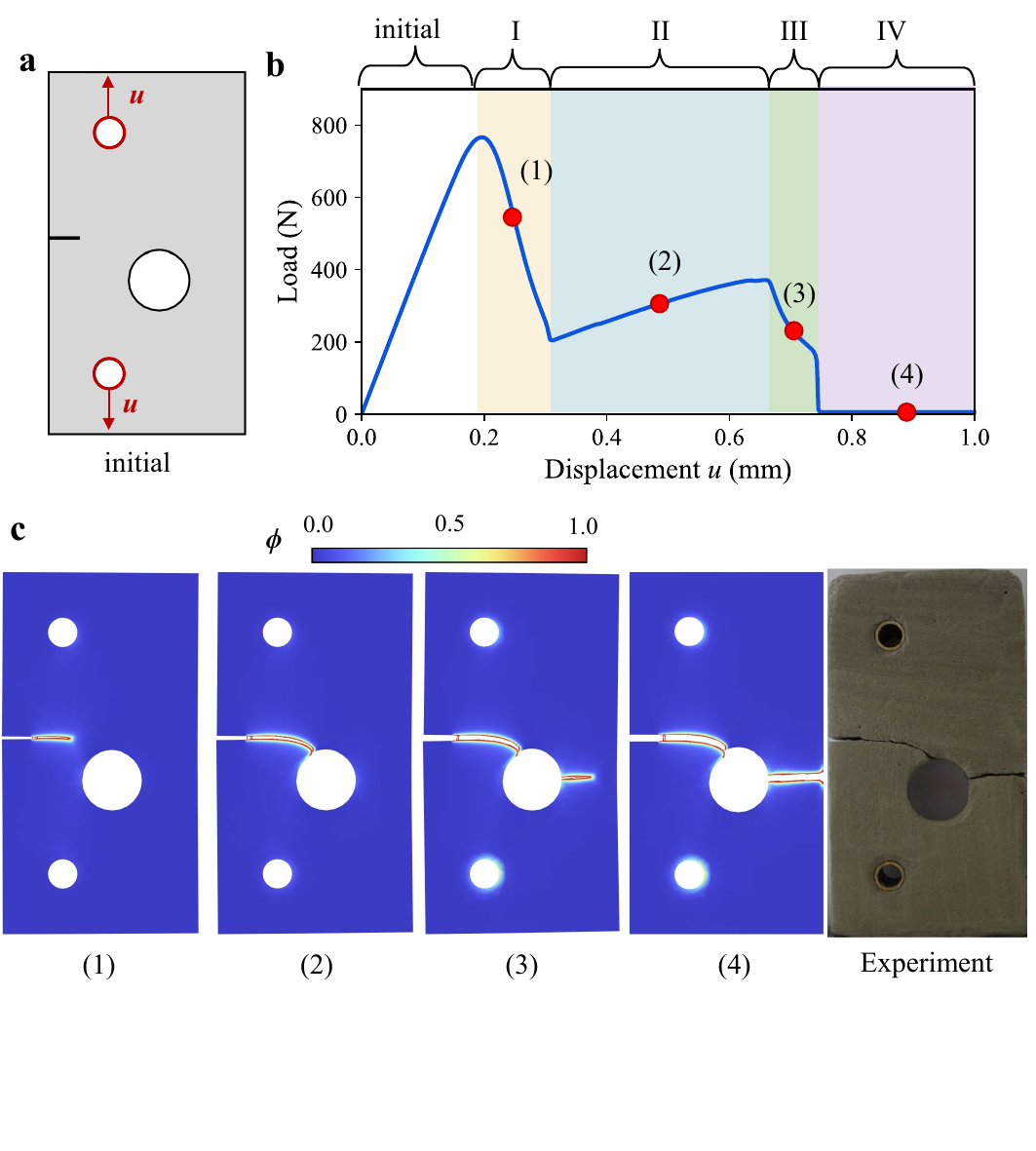}
    \caption{Crack propagation in the perforated plate under tensile loading.
(a) Geometry and boundary conditions.
(b) Load--displacement response and the four characteristic stages of crack evolution.
(c) Phase-field distributions at the selected loading states (1)--(4), together with the experimentally observed crack path. Elements deleted at \(\phi > \phi_{\mathrm{del}}\) are not shown.
The experimental image in (c) is reproduced from Ambati et al.~\cite{ambati2015review}, with permission from Springer Nature; copyright \copyright~Springer-Verlag Berlin Heidelberg 2014.
Additional rendering of this scenario is provided in Supplementary {Movie S1}.
}
    \label{fig:perforatedPlate}
\end{figure}

A perforated mortar plate subjected to displacement-controlled tension is considered to examine mixed-mode crack propagation under finite deformation.
The setup follows the benchmark based on the experiments of Ambati et al.~\cite{ambati2015review}.
As illustrated in Fig.~\ref{fig:perforatedPlate}, the plate has dimensions $65\times120~\mathrm{mm^2}$ and a thickness of $1~\mathrm{mm}$, with an edge notch, an eccentrically located central hole, and two smaller holes through which vertical displacements are applied.
The material and fracture parameters  are $E=6~\mathrm{GPa}$, $\nu=0.22$, $G_c=2280~\mathrm{J/m^2}$, and $\ell_0=0.25~\mathrm{mm}$~\cite{ambati2015review}.

Fig.~\ref{fig:perforatedPlate}(b) presents the load--displacement response, which can be divided into four stages.
After the initial peak, the crack propagates from the notch towards the central hole, causing a pronounced load reduction (Stage~I).
Once the crack reaches the hole, a temporarily stable configuration is established and the load increases again with a substantially reduced stiffness (Stage~II).
A second crack then initiates from the opposite side of the hole owing to the local stress concentration and propagates towards the right boundary (Stage~III), eventually separating the plate (Stage~IV).
The corresponding phase-field contours at the four marked loading points are shown in Fig.~\ref{fig:perforatedPlate}(c), and the final crack path agrees well with the experimental observation.
Moreover, the geometrically exact DDG formulation naturally accommodates the considerable changes in configuration throughout this process, highlighting its suitability for fracture problems involving geometric nonlinearity.

\subsection{Three-dimensional tearing of a plate}
\label{sec:three-dimensional-tearing}

\begin{figure}[t]
    \centering
    \includegraphics[
        width=0.75\linewidth,
        trim={0 0 0 0},
        clip
    ]{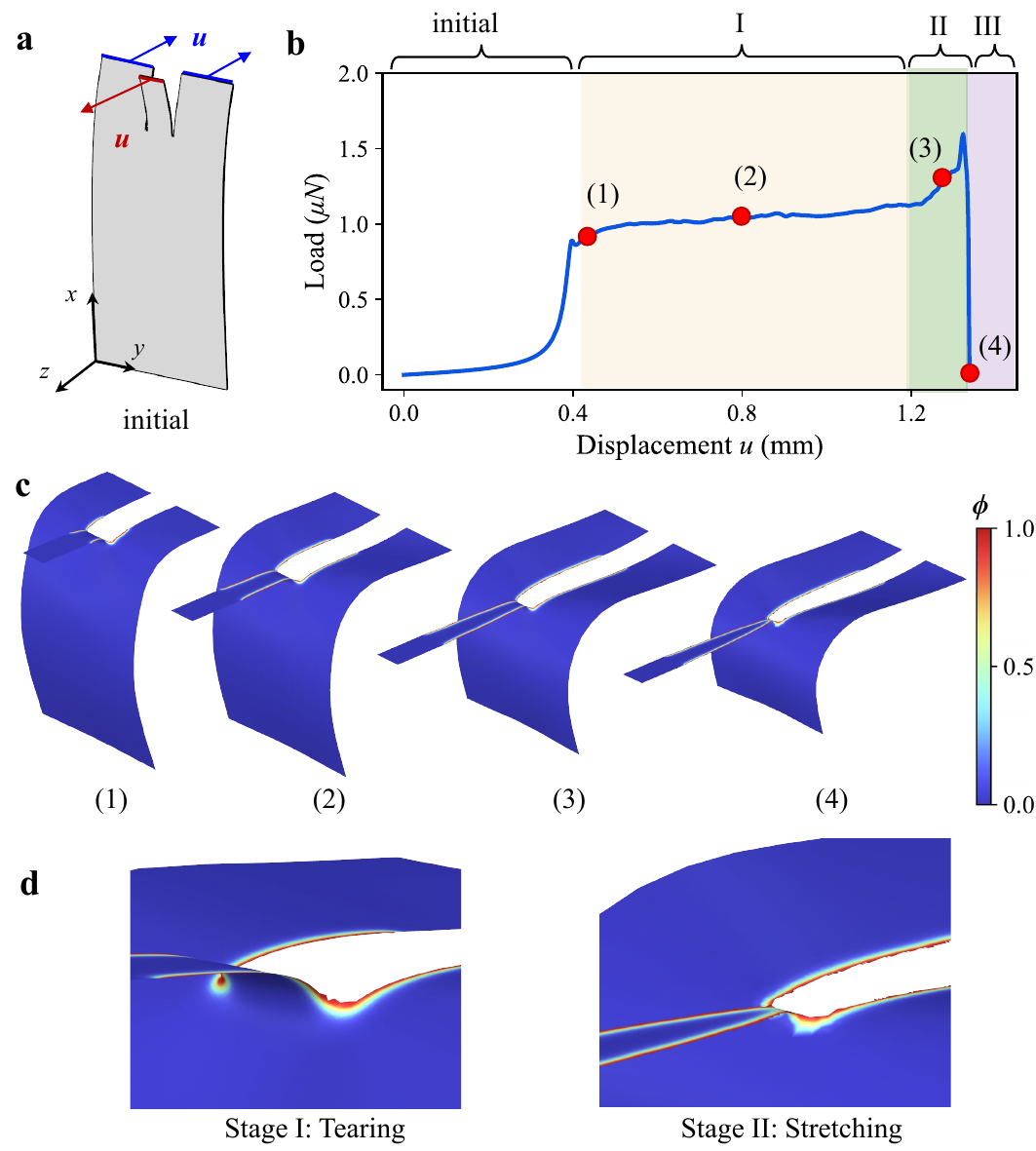}
    \caption{Three-dimensional tearing of a thin sheet.
(a) Geometry and boundary conditions.
(b) Load--displacement response and characteristic loading stages.
(c) Phase-field evolution at selected loading states.
(d) Crack evolution during Stage I (tearing) and Stage II (stretching).
Elements deleted at \(\phi > \phi_{\mathrm{del}}\) are not shown.
Additional rendering of this scenario is provided in Supplementary {Movie S2}.
}
    \label{fig:tearing_plate}
\end{figure}

A thin plate containing two initially parallel cracks is considered to demonstrate the capability of the proposed framework for three-dimensional tearing.
The configuration follows the trousers-type tearing test investigated by Bayart et al.~\cite{bayart2010tearing,bayart2011Finitedistance}, in which the central strip is pulled oppositely to the two outer strips, inducing substantial out-of-plane deformation and causing the two cracks to converge.
As shown in Fig.~\ref{fig:tearing_plate}(a), the plate has in-plane dimensions of $0.5\times1~\mathrm{mm^2}$ and a thickness of $h=5\times10^{-3}~\mathrm{mm}$.
The material and fracture parameters, adopted from Li et al.~\cite{li2016variational}, are $E=1~\mathrm{GPa}$, $\nu=0.3$, $G_c=20~\mathrm{J/m^2}$, and $\ell_0=7.5\times10^{-3}~\mathrm{mm}$.

The load--displacement response and the corresponding phase-field evolution are presented in Fig.~\ref{fig:tearing_plate}(b,c).
Oppositely directed displacements are applied to the central and outer strips, driving continuous crack propagation under an approximately constant tearing force during Stage~I.
This stable tearing regime persists over most of the loading process.
As observed by Bayart et al.~\cite{bayart2011Finitedistance}, the two cracks do not propagate along parallel paths but gradually curve towards each other and eventually coalesce, owing to the coupled bending and stretching of the sheet near the crack tips.
As shown in Fig.~\ref{fig:tearing_plate}(d), crack propagation during Stage~I is dominated by tearing and is locally aligned with the loading direction, i.e., the $z$-direction.
As the crack tips approach each other, the propagation direction rotates to become approximately perpendicular to the loading direction, and the narrowing ligament is increasingly subjected to stretching, as shown in Fig.~\ref{fig:tearing_plate}(e).
This transition produces the increase in load preceding final crack coalescence.

\begin{figure}[h]
    \centering
    \includegraphics[
        width=0.9\linewidth,
        trim={0 10cm 0 0},
        clip
    ]{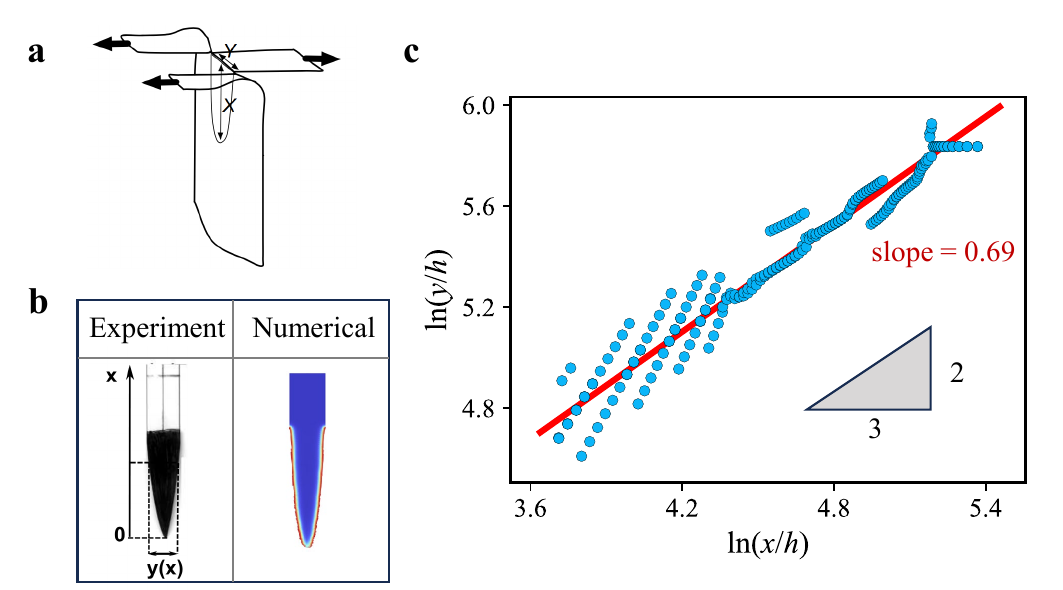}
    \caption{Formation of a tongue-like inner flap during crack convergence.
(a) Definition of the longitudinal and transverse coordinates.
(b) Comparison between the experimental and numerical strip profiles.
(c) Log--log representation of the numerical boundary and the fitted power-law scaling.
The experimental profile in (b) is adapted from Bayart et al.~\cite{bayart2011Finitedistance}, with permission from the American Physical Society.}
    \label{fig:tearing_profile}
\end{figure}

To quantify the converging crack paths, the detached central strip is extracted from the numerical model, and its topological boundary is compared with the experimental profile reported by Bayart et al.~\cite{bayart2011Finitedistance}.
As illustrated in Fig.~\ref{fig:tearing_profile}(a,b), the crack-coalescence point is used as the reference point, while $x$ and $y$ denote the longitudinal and transverse dimensions of the strip measured parallel and perpendicular to the initial cracks, respectively.
The resulting boundary is described by the power law $y\sim x^{\alpha}$, with the theoretical scaling predicting $y\sim x^{2/3}$.
The nearly linear relationship between $\ln(y/h)$ and $\ln(x/h)$ in Fig.~\ref{fig:tearing_profile}(c) confirms that the simulated boundary exhibits a power-law correlation between $x$ and $y$.
Linear regression gives $\alpha=0.69$, which agrees closely with the theoretical scaling $\alpha=2/3$ and lies within the experimental range $0.64\pm0.06$ reported by Bayart et al.~\cite{bayart2011Finitedistance}.

\begin{figure}[h]
    \centering
    \includegraphics[
        width=0.9\linewidth,
        trim={0 12cm 0 0},
        clip
    ]{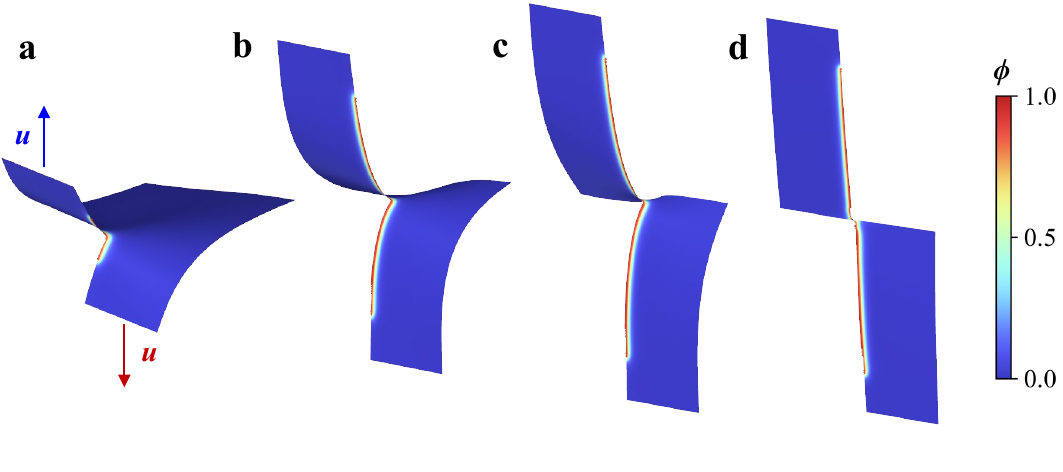}
    \caption{Straight crack propagation in a single-crack tearing test.
(a)--(d) Phase-field evolution and deformed configurations during the loading process.
Elements deleted at \(\phi > \phi_{\mathrm{del}}\) are not shown.
Additional rendering of this scenario is provided in Supplementary {Movie S3}.
}
    \label{fig:single_crack_tearing}
\end{figure}

As a supplementary example of three-dimensional tearing, a plate containing a single initial crack is also examined.
In contrast to the converging crack paths observed in the trousers-type test, the crack advances along an approximately straight path aligned with its initial direction throughout the loading sequence shown in Fig.~\ref{fig:single_crack_tearing}(a--d).
These two geometrically similar tests therefore produce distinctly different propagation patterns, indicating that crack growth in flexible structures is strongly influenced by deformation-induced changes in the stress state near the crack tip.
The geometrically exact nature of DDG makes the proposed framework particularly well suited to phase-field fracture simulations of shells undergoing finite deformation.

\subsection{Fracture of an island--bridge structure in flexible electronics}
\label{sec:island_bridge}

\begin{figure}[h]
    \centering
    \includegraphics[
        width=0.9\linewidth,
        trim={0 3cm 0 0},
        clip
    ]{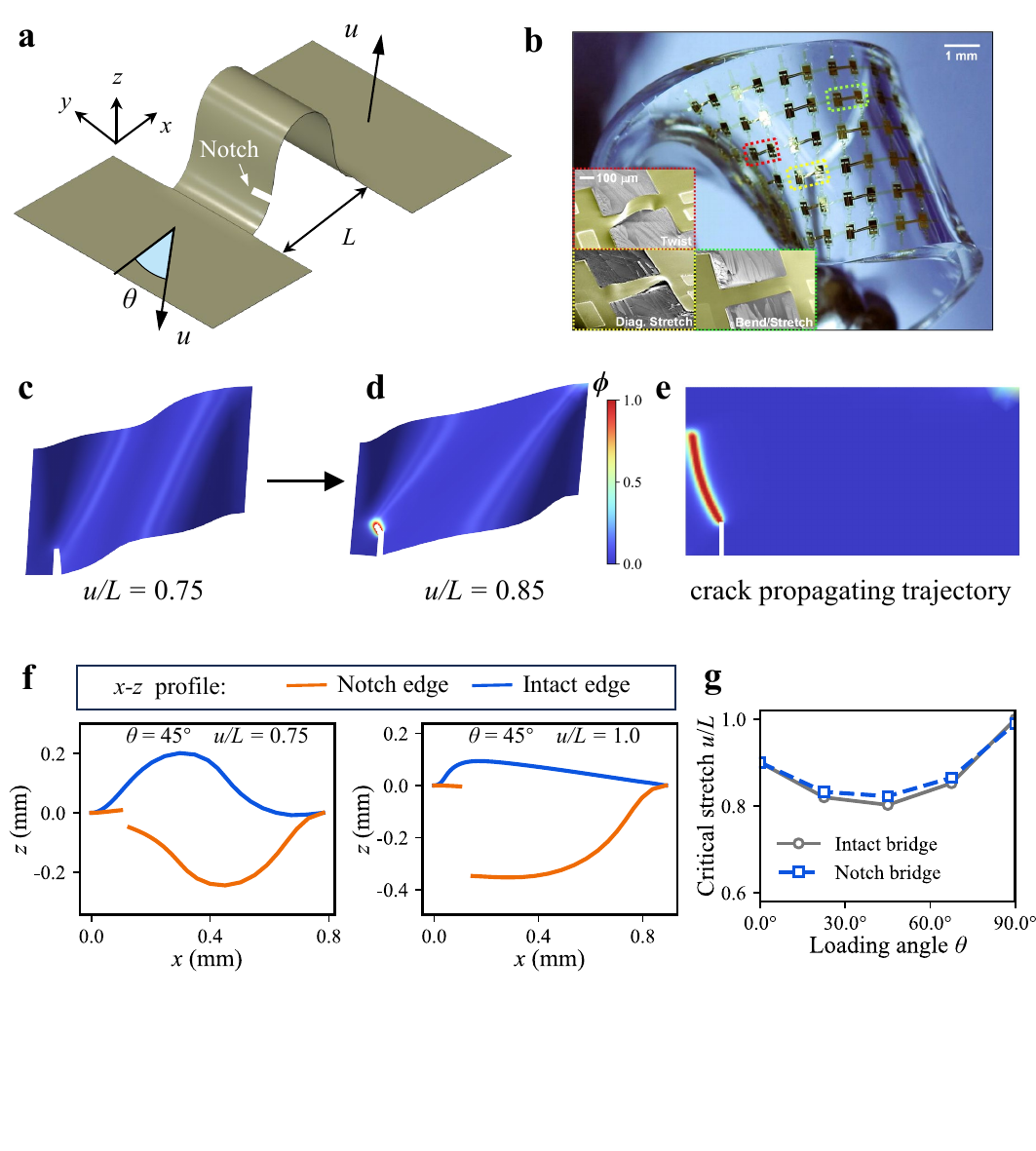}
    \caption{Fracture of a stretchable island--bridge structure.
(a) Geometry and loading configuration of the island--bridge structure.
(b) Deformed configurations observed experimentally under different loading directions.
(c) Diagonal-stretching configuration at $u/L=0.75$ for $\theta=45^\circ$.
(d) Onset and subsequent propagation of the crack from the notch tip. Elements deleted at \(\phi > \phi_{\mathrm{del}}\) are not shown.
(e) Crack path after complete propagation across the bridge.
(f) Projected boundary profiles before and after crack propagation.
(g) Critical stretch $u/L=0.85$ associated with the onset of crack growth as a function of the loading angle $\theta$ for the intact and notched bridges.
The experimental image in (b) is reproduced from Kim et al.~\cite{kim2008Materials}, with permission.
Additional rendering of this scenario is provided in Supplementary {Movie S4}.
}
    \label{fig:island_bridge}
\end{figure}

Island--bridge architectures are widely used in flexible electronics to provide stretchability through structural deformation.
As illustrated in Fig.~\ref{fig:island_bridge}(a), a typical architecture consists of two flat rectangular islands connected by a compliant bridge, which accommodates the imposed deformation while protecting the electronic components supported on the islands.
The bridge consequently undergoes large deformation and carries the associated stresses, potentially leading to fracture.
This produces a strong interaction between the evolving structural configuration and the crack-tip stress state.
The proposed framework is applied to trace the response of the bridge from stable large deformation to dynamic snap-through~\cite{huang2020Shear} and to examine how the accompanying changes in structural configuration influence the onset and subsequent evolution of crack growth.

Experimental observations by Kim et al.~\cite{kim2008Materials}, reproduced in Fig.~\ref{fig:island_bridge}(b), show that the bridge can exhibit bend/stretch, diagonal-stretching, and twisting modes depending on the loading direction.
These characteristic modes are reproduced using the proposed framework by varying the angle $\theta$ between the relative displacement of the islands.
Previous studies have shown that, for $\theta\neq0$, the transition from the initial configuration in Fig.~\ref{fig:island_bridge}(a) to the twisted configuration in Fig.~\ref{fig:island_bridge}(b) is associated with a supercritical pitchfork bifurcation~\cite{huang2020Shear}.
In the present framework, the dynamic formulation in Eq.~\eqref{eq:discrete-mechanical-equilibrium} captures the resulting snap-through during the transition between equilibrium branches.
The rectangular bridge has an in-plane dimensions of $0.5\times1~\mathrm{mm^2}$ and a thickness of $h=5\times10^{-3}~\mathrm{mm}$.
The material and fracture parameters, are $E=0.21~\mathrm{GPa}$, $\nu=0.3$, $G_c=27~\mathrm{J/m^2}$, and $\ell_0=7.5\times10^{-3}~\mathrm{mm}$. 

To investigate fracture under stretching, a notch of length
$0.1~\mathrm{mm}$ and width $0.015~\mathrm{mm}$ is introduced at the
bridge edge, with its center located $0.1~\mathrm{mm}$ from the left
boundary.
The loading angle is first fixed at $\theta=45^\circ$, Fig.~\ref{fig:island_bridge}(c)--(e) shows the resulting deformation
and crack evolution.
At $u/L=0.75$, the bridge exhibits the characteristic diagonal-stretching configuration observed experimentally in Fig.~\ref{fig:island_bridge}(b).
As the applied displacement increases, crack propagation initiates from
the notch tip at approximately $u/L=0.85$, as shown in
Fig.~\ref{fig:island_bridge}(d).
The crack subsequently propagates across the bridge and reaches the
left boundary at $u/L=1.0$.
The resulting trajectory in Fig.~\ref{fig:island_bridge}(e) is
approximately transverse to the bridge diagonal, consistent with crack
growth driven by the tensile deformation concentrated along the diagonal
stretching band~\cite{huang2020Shear}.

The influence of the crack on the large-deformation response is further
examined through the projected boundary profiles in
Fig.~\ref{fig:island_bridge}(f).
At $u/L=0.75$, opening of the notch separates the notched edge and produces a pronounced tearing-like configuration, while the opposite edge remains continuous.
This geometric separation localizes the deformation near the notch tip, giving rise to the stress concentration that subsequently drives the onset of crack growth.
After the crack has fully propagated at $u/L=1.0$, the release of the boundary constraint allows much of the bending deformation to relax, resulting in an overall reduction in the out-of-plane curvature of the bridge.
Meanwhile, fracture alters the initially center-symmetric deformation pattern, giving rise to the asymmetric boundary profiles observed after fracture.

The influence of the notch on the stretchability of the bridge is quantified in Fig.~\ref{fig:island_bridge}(g).
The critical stretch ratio $u/L$ associated with the onset of crack growth is evaluated by varying the loading angle $\theta$ from $0^\circ$ to $90^\circ$, and the results for the intact bridge and the bridge containing the notch shown in Fig.~\ref{fig:island_bridge}(a) are compared.
In both cases, the critical stretch ratio is higher near $\theta=0^\circ$ and $90^\circ$ and reaches its minimum near $\theta=45^\circ$, indicating greater stretchability under the pure tensile and shear loading modes than under intermediate loading directions.
Introducing the notch results in a slight increase in the critical stretch ratio.
Although the notch promotes local stress concentration, its opening provides an additional deformation mechanism that accommodates part of the imposed extension and thereby delays the onset of crack growth.
For this reason, bridges with cutout geometries have attracted considerable attention owing to their enhanced stretchability~\cite{zhang2013Bucklingb}.

\section{Conclusions}

This work developed a DDG--phase-field framework for simulating fracture in highly deformable thin structures.
The shell midsurface is represented by a triangular mesh, on which membrane stretching and bending are formulated through discrete counterparts of the first and second fundamental forms.
Crack evolution is described by a phase field defined on the same discrete surface, providing a geometrically consistent coupling between structural deformation and fracture.
The resulting mechanical and phase-field problems are solved using a staggered scheme, with crack irreversibility enforced through an active-set method.
A degradation--deletion procedure was further introduced to remove fully failed elements and reconstruct the DDG topology, thereby permitting complete crack opening and substantial post-fracture reconfiguration.

The numerical examples demonstrated the capability of the framework across deformation and fracture regimes of increasing complexity.
For the pre-cracked plate, the predicted crack path and load--displacement response agreed closely with reference finite element results. 
The perforated plate reproduced the experimentally observed sequence of crack propagation, temporary arrest at the central hole, reinitiation, and final separation under considerable changes in configuration.
In the three-dimensional tearing tests, the formulation captured the coupling between out-of-plane deformation and crack-path evolution: two initially parallel cracks gradually converged and coalesced, whereas a single crack remained approximately straight.
These results show that changes in structural geometry can substantially modify the crack-tip stress state and hence the direction and interaction of propagating cracks.
Finally, the island--bridge example demonstrated the capability of the proposed framework to resolve the coupled effects of large deformation, structural instability, and fracture in a representative flexible system.
The simulations captured loading-dependent deformation modes and snap-through, and further showed that crack propagation are strongly governed by the evolving structural configuration and its interaction with pre-existing defects.

These observations highlight a broader class of problems in which geometric reconfiguration is not merely a consequence of loading, but an essential part of structural functionality and failure.
The proposed framework therefore provides a potential computational basis for studying damage and fracture in highly deformable systems such as stretchable electronics, soft robotic structures, compliant mechanisms, and reconfigurable or deployable thin structures, where large deformation, instability, and fracture may evolve in a strongly coupled manner.

\section*{Acknowledgments}
B.Z. acknowledges financial support from China Scholarship Council.
B.W. acknowledge the support from the National Natural Science Foundation of China (No. 12672002, 12432002, 12402002, 62388101), from the Guangdong Basic and Applied Basic Research Foundation (No. 2024A1515010767).
W.H. acknowledges the start-up funding from Newcastle University, UK.

\appendix
\renewcommand{\thesection}{Appendix \Alph{section}}

\section{Video}
\label{sec:AppendixA}

We provide four videos as supplementary material to illustrate our  results.

\bibliographystyle{elsarticle-num-names}
\bibliography{Bibliography.bib}

\end{document}